\documentclass[10pt,journal,compsoc]{IEEEtran}
\usepackage{latexsym}
\usepackage{graphicx}
\usepackage{amsfonts,amssymb,amsmath}
\usepackage{hyperref}
\usepackage{array}
\usepackage{multirow}
\usepackage{xcolor}
\usepackage{subfigure}
\usepackage{rotating}
\usepackage{makecell}
\usepackage{ulem}
\begin{document}

\title{Mitigating Gradient-based Adversarial Attacks\\ via Denoising and Compression}

\author{Rehana Mahfuz, \textit{Student Member, IEEE}, Rajeev Sahay, \textit{Student Member, IEEE},\\ Aly El Gamal, \textit{Senior Member, IEEE}
\thanks{The authors are with the School of Electrical and Computer Engineering, Purdue University, West Lafayette, IN 47906 USA (e-mail: rmahfuz@purdue.edu, sahayr@purdue.edu, elgamala@purdue.edu).

Rehana and Rajeev have equally contributed to this work.

This work has been presented in part at the 2019 Annual Conference on Information Sceinces and Systems (CISS) \cite{ciss_cascade_paper}.
}}

\maketitle


\begin{abstract}Gradient-based adversarial attacks on deep neural networks pose a serious threat, since they can be deployed by adding imperceptible perturbations to the test data of any network, and the risk they introduce cannot be assessed through the network's original training performance. Denoising and dimensionality reduction are two distinct methods that have been independently investigated to combat such attacks. While denoising offers the ability to tailor the defense to the specific nature of the attack, dimensionality reduction offers the advantage of potentially removing previously unseen perturbations, along with reducing the training time of the network being defended. We propose strategies to combine the advantages of these two defense mechanisms. First, we propose the cascaded defense, which involves denoising followed by dimensionality reduction. To reduce the training time of the defense for a small trade-off in performance, we propose the hidden layer defense, which involves feeding the output of the encoder of a denoising autoencoder into the network. Further, we discuss how adaptive attacks against these defenses could become significantly weak when an alternative defense is used, or when no defense is used. In this light, we propose a new metric to evaluate a defense which measures the sensitivity of the adaptive attack to modifications in the defense. Finally, we present a guideline for building an ordered repertoire of defenses, a.k.a. a defense infrastructure, that adjusts to limited computational resources in presence of uncertainty about the attack strategy.

\end{abstract}



\section{INTRODUCTION}

\IEEEPARstart{A}{dversarial} attacks on deep neural networks pose a serious threat to systems that rely on the performance of these networks for critical applications. Evasion attacks are capable of destroying the performance of networks by introducing perturbation only to test data, requiring no changes to be made to the weights or hyperparameters of the network. Moreover, it has been shown that the attacker is typically able to make these perturbations imperceptible, which means that the presence of the attack is not obvious even when the test data is examined. Hence, there is a need to develop defense strategies for such attacks, keeping in mind the limitations of both the attacker and the defender. Most of the defense strategies proposed so far either demand excessive computation, such as adversarial training, or are vulnerable to adaptive attacks.

\subsection{Threat Model}
In this work, we restrict our attention to gradient-based adversarial evasion attacks, where the attacker adds a perturbation to the input samples at test time, based on gradient-based optimization that typically matches the nature of training optimization for deep neural networks where parameters updates are based on the gradient of a cost function. We refer to the neural network vulnerable to attacks as the \textit{victim network} or \textit{victim classifier}, and the system consisting of the neural network and any possible set of defenses as the \textit{victim system}. As mentioned earlier, the success of evasion attacks does not rely on manipulating any part of the victim system. However, the amount of information the attacker has about the victim system largely determines how successful the attack is. Having knowledge of the exact parameters of the victim system allows the attacker to generate a stronger attack than merely having limited knowledge of the nature of the victim system. For the victim network, these parameters may include the number, size and activations of layers in the network, the loss function, the optimizer, potentially employed regularization techniques, weight initialization strategy, and the randomly initialized weights. For the victim defense, important examples of such parameters include the number of bits in the bit depth reduction defense \cite{feature_squeezing}, the trained GAN in APE-GAN \cite{ape_gan_defense} and Defense-GAN \cite{defense_gan_defense}, the trained detector and reformer networks in MagNet \cite{magnet_defense}, and the range of sizes to resize to if mitigating through randomization \cite{mitigate_thru_rand_defense}. 
Apart from knowledge of the victim system, the other major determiner of the success of the attack is computational capacity. For instance, merely moving the data sample in the direction opposite to the gradient of the loss function results in a moderately strong attack \cite{fgs_attack}. However, a stronger attack is generated if the attacker sets up an objective to minimize the perturbation while satisfying the constraint of misclassification, and employs binary search to determine a good relative scale of the terms in the objective, uses gradient descent with multiple random starting points, and continues for a large number of iterations \cite{cw_attack}.

In the evaluation of considered defenses, we operate in a conservative threat model where we assume that the attacker has exact knowledge of the victim system, i.e., the attacker is allowed to not only probe the neural network with inputs as in \cite{query_mech}, but is given direct access to the architecture and weights of the trained neural network. Further, it is also given access to knowledge of all available defenses, if there are any. Such a transparent setting is commonly referred to as a \textit{white box setting}. Our setting deviates slightly from the strict white box setting by randomizing the choice of defense at test time, and potentially hiding knowledge of the exact choice from the attacker.

\section{Background}
\subsection{Attack algorithms}
There are many ways to generate gradient-based adversarial attacks against neural networks. Here, we describe four representative attack algorithms for a classification setting that we consider in this work. The Fast Gradient Sign (FGS) method attempts to maximize the loss function of the victim network by adding a perturbation to the input which is in the direction of the sign of the gradient of the loss function $J(\theta, x, y)$, where $\theta$ refers to the trained weights, $x$ is the input sample, and $y$ is its target label. From the original sample $x$, the perturbed sample $\tilde{x}$ is generated by adding a perturbation $\delta$, whose $l_2$ norm is restricted by $\epsilon$ to maintain imperceptibility. This perturbation is calculated as
\begin{equation}
\delta = \epsilon \frac{\Delta_xJ(\theta, x, y)}{||\Delta_xJ(\theta, x, y)||_2}.
\end{equation}
The Projected Gradient Descent (PGD) attack performs the same step followed by a projection, for multiple iterations. The perturbation to be added at step $i+1$ is
\begin{equation}
    \delta_{i+1} = P_\epsilon \bigg( \delta_i + \alpha* \frac{\Delta_xJ(\theta, x, y)}{||\Delta_xJ(\theta, x, y)||_2} \bigg),
\end{equation}
where \begin{equation}
    P_\epsilon(z) = \epsilon \frac{z}{max\{\epsilon, ||z||_2\}}.
\end{equation}
Here, $\epsilon$ is the limit on the $l_2$ norm of the total distortion, and $\alpha$ is the limit on the $l_2$ norm of the distortion at each iteration.

The Carlini Wagner (CW) attack attempts to find the optimal perturbation $\delta$ of the following problem:
\begin{equation}
    \min_{\delta} ||\delta||^2 + c.\max(Z_{C^*(x)}(x+\delta) - \max_{i \neq C^*(x)}\{Z_i(x+\delta)\}, 0),
\end{equation}
where $C^*(x)$ is the true label, the notation $Z_j(x')$ is used to refer to the output of the $j^{th}$ Softmax logit when the input is $x'$, and $c$ is a constant determined by binary search. Here, logits refer to the raw prediction probabilities of a neural network that are commonly fed into a softmax function. For image data, since the perturbed data sample $x+\delta$ has to lie in a certain range to be interpreted as a pixel, a change of variable is applied to $\delta$ to ensure that it lies between 0 and 1:
\begin{equation}
    \delta = \sigma(2w) - x.
\end{equation}
Gradient descent with multiple random starting points is used to solve the problem.

The Deepfool (DF) attack \cite{df_attack} attempts to move the input sample across the boundary of the closest separating hyperplane. This is performed for a certain number of iterations (generally 50), or until the classifier is found to misclassify the perturbed data. At every iteration $i$, by assuming the decision boundaries to be linear at the point closest to the data point $x_i$, a region $\tilde{P_i}$ is approximated such that $x_i$ would be correctly classified in that region:
\begin{equation}
\begin{split}
    \tilde{P_i} = \bigcap_{k = 1}^{|C|} \{x: f_k(x_i) + \nabla f_k(x_{i})^Tx \\ 
    \leq f_{C^*(x_0)}(x_i) + \nabla f_{C^*(x_0)}(x_i)^Tx\}.
\end{split}    
\end{equation}
Here, $f_k(x')$ is a real-valued output of classifier $f$ corresponding to class $k$ for an input $x'$. The decision of the classifier is given by
$\displaystyle\arg\max_k f_k(x')$

At iteration $i$, the perturbation $\delta_i$ is calculated as:
\begin{equation}
    \delta_i = \frac{|f_l(x_i) - f_{C^*(x_0)}(x_i)|}{||\nabla f_l(x_i) - \nabla f_{C^*(x_0)}(x_i)||_2^2}(\nabla f_l(x_i) - \nabla f_{C^*(x_0)}(x_i)),
\end{equation}

where $l$ is the index of the class closest to $x_0$, which is found as:
\begin{equation}
    l = arg min_{k \neq C^*(x_0)} \frac{|f_k(x_i) - f_{C^*(x_0)}(x_i)|}{||\nabla f_k(x_i) - \nabla f_{C^*(x_0)}(x_i)||_2^2}.
\end{equation}

This is a good strategy if the attacker has access to the victim classifier's exact weights, since it enables finding a very small effective perturbation. However, this attack typically does not work well if deployed on a different classifier, since it perturbs the input only as much as required for that specific classifier whose gradients are used to craft the attack.

\subsection{Defense strategies}
Strategies to defend neural networks against adversarial attacks may involve modifying the network itself, modifying the training process, or modifying the data that is input into the network.
Modifying the network may involve adding a new class to assign to adversarial examples \cite{null_labeling}, using a special activation function such as bounded ReLU \cite{bounded_relu}, controlling the sensitivity to input perturbations by penalizing the degree of change in the output for a small change in the input via gradient regularization \cite{gradient_regularization}, or by controlling a Lipschitz constant at each layer \cite{parseval_net}. It is worth noting that making modifications to the network can be tedious, especially when having to use pre-trained networks for large datasets.

Important examples for modifying the training process include network distillation \cite{distillation} and adversarial training \cite{adv_train}. Network distillation  is the process of training the network with soft labels which are generated by a different network with a modified softmax layer as follows, where $T$ is a temperature set to a value greater than $1$ while training:
\begin{equation}
    f_k(x) = \frac{e^{\frac{Z_k(x)}{T}}}{\sum_{l=1}^{|C|} e^{\frac{Z_{l}(x)}{T}}}.
\end{equation}
Note that defensive network distillation has been effectively compromised by the CW attack \cite{cw_attack}. Adversarial training is the process of training a network to correctly classify adversarial examples in addition to classifying clean samples. One variant is cascade adversarial training \cite{cascade_adv_train}, where the victim network is trained with adversarial examples generated using a different network, in addition to adversarial examples generated using the same network. This is very computationally expensive, as it requires the entire network to be trained with more data. Also, this is likely to perform best only if the attacker happens to use the same type of attack that the network is trained to overcome.

Example mechanisms for modifying the data input into the network include randomization, compression, and denoising. For randomization, \cite{mitigate_thru_rand_defense} suggested random rescaling of test images followed by random zero padding as a defense. For compression, which is also known as dimensionality reduction, methods like Principal Components Analysis \cite{bhagoji_pca_defense} or an autoencoder \cite{ciss_cascade_paper} can be employed regardless of data type, while methods like JPEG compression \cite{jpg_das} and feature squeezing \cite{feature_squeezing} have been designed and tested only with images. An advantage of dimensionality reduction is that the victim classifier processes a smaller input dimension, which reduces its training time significantly. It is worth noting that if a pre-trained classifier already exists, then re-training the first layer or the first few layers is generally sufficient to accommodate the reduced dimension of the input. For denoising, a denoising autoencoder may be trained to remove the perturbation, as in \cite{ciss_cascade_paper}, or a convolutional denoising autoencoder specialized for color images, that learns the structure of noise to be subtracted, as in \cite{dunet}, or the denoiser in \cite{magnet_defense}. Also, a Generative Adversarial Network (GAN) was used for denoising in APE-GAN \cite{ape_gan_defense}. Further, Defense-GAN \cite{defense_gan_defense} used a GAN to generate a clean image as close as possible to a given test image. We note that there are several serious issues associated with having to use GANs including an intricate training process that is computationally demanding and highly prone to generalization errors.

Once the attacker knows about the defense being used, it is common for the attacker to custom-tailor an attack to overcome that specific defense, as illustrated in \cite{obf_grad}. Such an attack is known as an \textit{adaptive attack}. To generate an adaptive attack, the attacker considers the system including the classifier and the defense as a whole, and generates an attack against that. In addition to considering adaptive attacks to overcome one defense, we also consider in this work adaptive attacks to simultaneously overcome multiple defenses.

\section {Contributions}

Dimensionality reduction may be viewed as a general strategy to filter out noise, as it is not tailored to a particular attack type. On the other hand, explicit denoising strategies can be custom-tailored to the nature of the attack, if that is known. However, it is typically hard to predict the nature of the attack, which also poses a major limitation for adversarial training. In an effort to combat adversarial perturbations with the limited knowledge we may have about the anticipated attack, we combine the custom-tailored noise removal method of denoising with the general noise removal method of dimensionality reduction. The contributions of this work can be summarized as follows:
\begin{enumerate}
    \item \textbf{Individual defenses of modular nature:} 
    We propose two standalone defenses, which, if implemented individually, do not necessarily require re-training of the entire victim neural network or any other major modifications. Each defense is easy to add to and remove from the system, and may be used in combination with other defenses.
    \newline
    \item \textbf{Defenses requiring limited computational resources:} 
    Using either of our newly proposed defenses can reduce the training time of the network by as much as 42\%. Moreover, if the network is already trained, then our defense can be applied with re-training of only the first layer. 
    \newline
    \item \textbf{Guideline to overcome adaptive attacks:}
    In recognition of the fact that each of our defenses individually fails in a strict white box scenario when the attacker uses an adaptive attack, we provide insight into building a defense infrastructure which confuses the attacker about which defense from a repertoire of defenses is being used. This infrastructure takes advantage of the fact that an adaptive attack designed to circumvent a specific defense becomes weak when a different defense is used.
    We provide a guideline to building such an infrastructure that takes into consideration any trained networks that may already be available, hence prudently minimizing the amount of training required.
    Further, we have decomposed the building process into steps, such that each step achieves the complete addition of one or more defenses to the repertoire, while ensuring that termination of the process at the end of a step caused by limited computational resources would result in the availability of at least a bare minimum number of defenses to choose from.
    \newline
    \item \textbf{Metric to evaluate defenses:}
    Our results reveal a metric to evaluate defenses which takes into account the vulnerability of an adaptive attack to failure if that exact defense is not used, either because a different defense mechanism is used, or because changes are introduced to the same defense.
    \newline

\end{enumerate}

\section{Procedure}

\begin{figure}[t]
  \centering 
  \subfigure[DAE defense.]{\label{subfig: dae}  \includegraphics[width=0.45\textwidth]{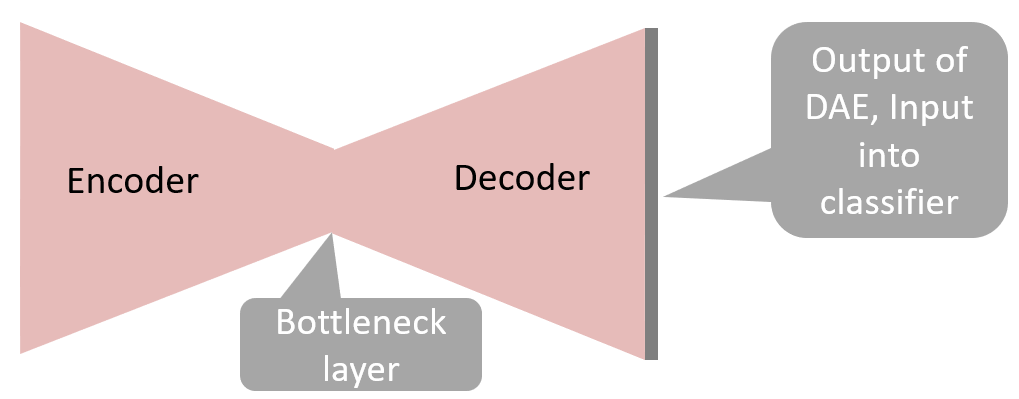}}%
  \hskip 0.5truein
  \subfigure[Cascaded defense.]{\label{subfig: cascade}\includegraphics[width=0.5\textwidth]{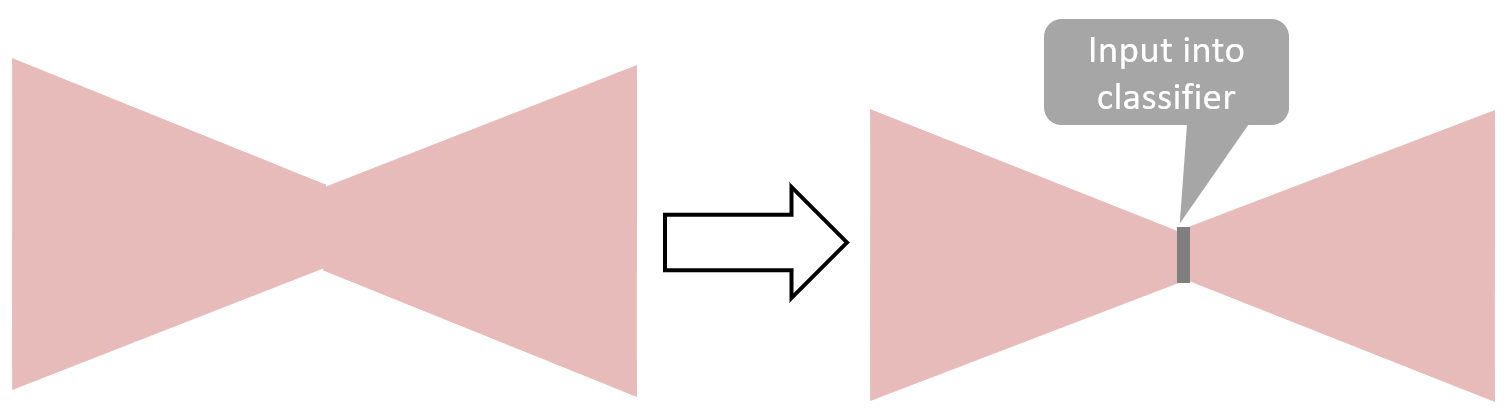}}
  \hskip 0.5truein
  \subfigure[Hidden Layer defense.]{\label{subfig: hidden_layer}\includegraphics[width=0.3\textwidth]{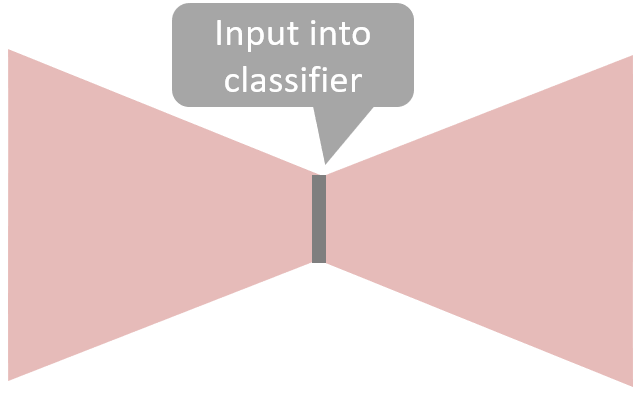}}
  \caption{Our three considered defenses that rely on DAEs.}
  \label{fi:two-parts}
\end{figure}

While it is possible to determine the nature of adversarial perturbations to a certain extent, a defender should not rely on strong assumptions regarding adversarial attacks. Hence, our initial intuition was to perform explicit denoising by taking advantage of possibly imperfect knowledge about potential attacks, and then introduce the capability to remove unseen perturbations using compression. 
An important advantage of compressing the input is that it reduces the training time of the original (victim) classifier. If there is no pre-trained classifier available, this advantage is obvious. When there is a pre-trained classifier available, all that is required to obtain this new classifier to handle compressed input is re-training of the first layer, modified to have reduced input dimension. Both classifiers, with and without reduced input dimension, eventually have to be trained when the goal is to build robustness against adaptive attacks, as explained later in Section \ref{subsec: guide}, and illustrated in Figure \ref{fig: tree}.

To perform denoising, we use a Denoising Autoencoder (DAE), denoted as $H(x)$, which consists of an encoder $p(x) = x'$: ${\rm I\!R}^n \rightarrow {\rm I\!R}^k$, that compresses the possibly perturbed $n$-dimensional input $x$ into a $k$-dimensional representation $x'$ ($k<n$), and a decoder $q(x') =\hat{x}$: ${\rm I\!R}^k \rightarrow {\rm I\!R}^n$, which reconstructs an approximation $\hat{x}$ of the original $n$-dimensional input $x$ from the compressed $k$-dimensional representation $x'$. This is illustrated in Figure \ref{subfig: dae}. The parameters of both the encoder and decoder are trainable. The output of the encoder is also called the output of the bottleneck layer of the autoencoder. We train the DAE to reconstruct a cleaned version of the input data, regardless of whether the input is clean or corrupted. This is done by minimizing the Mean Squared Error (MSE) between the clean target sample $\bar{x}$ and the generated output $q(p(x))$:

\begin{equation}\label{dae} 
    H(x) = \underset{p, q} {\text{minimize}} \frac{1}{n} \sum_{i=1}^{n}(\bar{x_i} - q(p(x_i)))^2.
\end{equation}

To perform dimensionality reduction, we again use an autoencoder, henceforth referred to as the \textit{compression autoencoder}. The output of its bottleneck layer is used as the compressed representation. This autoencoder is trained by setting the target output to be the same as the input, hence motivating the autoencoder to propagate the same information throughout the network without making any changes to it. However, since this information needs to be squeezed through a bottleneck layer of much lower dimension, the autoencoder strives to represent the same information using lesser dimensions, hence generating the compressed representation.

To implement our first defense that combines denoising and dimensionality reduction, which we call the \textbf{\textit{cascaded defense}}, the output of the DAE is fed into the compression autoencoder, as shown in Figure \ref{subfig: cascade}. The compressed representation from the bottleneck layer of this autoencoder is then fed into an AE-reduced classifier. An AE-reduced classifier is a neural network which has been trained to classify an input sample that is first compressed by the compression autoencoder. Such an AE-reduced classifier can be obtained from the original classifier in one of two ways. The first method involves reducing the size of the input layer of the original classifier to match the size of the compressed data representation. Even though such an AE-reduced classifier takes lesser time to train than the original classifier because the number of trainable parameters is reduced due to a reduction in the input size, this method requires us to train the entire network again, which is feasible when the network is small. However, for more complex networks, re-training may be expensive. Hence, our second method is to simply add an upsampling layer after the input layer of reduced size, so that we have to re-train only the first layer, and not the remaining layers. We re-use weights of all layers except the first, which significantly reduces training time. 
To obtain the AE-reduced classifier, we refer to the first method as the \textit{re-training method}, and to the second method as the \textit{upsampling method}.

We observe that the cascaded defense involves compression by the encoder of the DAE, followed by decompression by the decoder of the DAE, followed again by compression by the encoder of the compression autoencoder. To eliminate this redundancy, we attempted to use the output of the encoder of the DAE as the data that is fed into a DAE-reduced classifier. A DAE-reduced classifier is a neural network which has been trained to classify data that is compressed by the encoder of the DAE. 
Such a method of defending the possibly corrupted data is termed as the \textbf{\textit{hidden layer defense}}, also abbreviated as the \textit{HL defense}. This defense is illustrated in Figure \ref{subfig: hidden_layer}. The advantage of this method is that it eliminates the need to train the compression autoencoder. The \emph{re-training} and \emph{upsampling} methods of obtaining the DAE-reduced classifier are similar to the methods of obtaining the AE-reduced classifier.

\subsection{Implementation Details} \label{subsec: impl_details}

We used the MNIST-Digit, Fashion-MNIST and CIFAR-10 datasets. The first two datasets contain 28x28 pixel grayscale images of handwritten digits and fashion items, respectively. Each of these two datasets has 60,000 training images and 10,000 test images, belonging to one of ten classes. In the MNIST-Digit dataset, these classes correspond to the ten digits, while in the Fashion-MNIST dataset, these classes correspond to ten distinct fashion items such as shirt, trouser, sandal etc. The CIFAR-10 dataset consists of 50,000 training samples and 10,000 test samples of 32x32x3 pixel color images, belonging to one of ten categories such as airplane, cat, horse etc. 

A \emph{victim} deep neural network is used for input sample classification. For the first two datasets, the victim network is a Convolutional Neural Network (CNN) where the first convolutional layer has 32 3x3 convolution filters with ReLU activation, and the second convolutional layer has 64 such filters with the same activation\footnote{Code has been accepted for publication at: \url{https://codeocean.com/capsule/7548435/tree/v1}}. These layers are followed by a softmax output layer. Such a CNN achieves an accuracy of 98.61\% for the MNIST-Digit dataset and 93.32\% for the Fashion-MNIST dataset. Both CNNs were trained for 20 epochs with a batch size of 200, categorical crossentropy loss and Adam optimizer, which used a learning rate of 0.001 and exponential decay rates of 0.9 and 0.999 for the first and second gradient moments, respectively. The architecture of the victim CNN used for the CIFAR-10 classification task is shown in Table \ref{table:cifar_victim_cnn_arch}. The accuracy of this classifier is 90.44\%. To obtain the AE-reduced and DAE-reduced classifiers, the re-training method was used for the MNIST-Digit and Fashion-MNIST datasets, while the upsampling method was used for the CIFAR-10 dataset.

Additional CNNs with slightly varying architectures were also trained for the CIFAR-10 classification task. To obtain these variations, an extra (Conv 3x3x$z$, ELU, BatchNorm) sequence is added after the second, fourth, sixth, or eighth such sequence, where $z$ is the number of filters in the convolutional layer preceding the added convolutional layer. They achieve accuracies of 90.93\%, 90.33\%, 90.18\% and 89.85\%. The gradients of the first three of these variant CNNs are used to generate attacks to train the DAE.
The fourth such modified CNN is used as the adversary's CNN. 
Note that ELU refers to Exponential Linear Unit activation and BatchNorm refers to Batch Normalization. All of the CNNs for the CIFAR-10 classification task were trained for 225 epochs with a batch size of 64, categorical crossentropy loss and the RMSProp optimizer which starts with an initial learning rate of 0.001 with a decaying factor of $10^{-6}$, and the hyperparameter $\rho$ controlling the forget rate in the moving average of squared gradients is set to 0.9 (see \cite{dl-book}, Algorithm 8.5 for more details). Data augmentation was performed prior to feeding the CIFAR-10 data into the CNN, through rotations of up to 15 degrees, width/height shifts of up to 10\% of the original, and horizontal flips.

\begin{table}[]
 \caption{CIFAR-10 victim CNN architecture.}
 \begin{center}
 \begin{tabular}{ p{8cm} }
 \hline
 (Conv 3x3x32, ELU, BatchNorm)x2\\
 Max Pool 2x2, Dropout (rate = 0.2)\\
 (Conv 3x3x64, ELU, BatchNorm)x2\\
 Max Pool 2x2, Dropout (rate = 0.3)\\
 (Conv 3x3x128, ELU, BatchNorm)x2\\
 Max Pool 2x2, Dropout (rate = 0.4)\\
 (Conv 3x3x128, ELU, BatchNorm)x2\\
 Max Pool 2x2, Dropout (rate = 0.4)\\
 Softmax (10 classes)\\
 \hline
 \end{tabular}
 \end{center}
 \label{table:cifar_victim_cnn_arch}
 \end{table}

For the first two datasets, the DAE has architecture FC-784-256-128-81-128-256-784. Here, the architecture FC-$n_1-n2-...-n_k$ describes a neural network, with fully connected layers and $n_1$ neurons in its first layer, $n_2$ neurons in its second layer, and so on, till the $k^{th}$ layer, which has $n_k$ neurons. None of the layers has any activation except the final layer, which has sigmoid activation. This DAE was trained for 150 epochs with a batch size of 200, Mean Squared Error (MSE) loss and Adam optimizer which uses a learning rate of 0.001 and exponential decay rates of 0.9 and 0.999 for the first and second gradient moments, respectively. Regarding the choice of the attack type, we trained the DAE with the CW, DF and FGS attacks which were generated using gradients of the victim network. The FGS attack used an $l_2$ norm of 1.5. The architecture of the CIFAR-10 DAE is shown in Table \ref{table:cifar_dae_arch}. To make the CIFAR-10 DAE more resilient, it was trained not only with perturbations generated using gradients of the victim CNN, but also with perturbations generated using gradients of the three CNNs, described above, with modified architectures. It was trained for 150 epochs with a batch size of 256, with MSE loss and Adam optimizer, which uses a learning rate of 0.001 and exponential decay rates of 0.9 and 0.999 for the first and second gradient moments, respectively.

\begin{table}[h]
\caption{CIFAR-10 DAE architecture.}
\begin{center}
 \begin{tabular}{ p{8cm} }
 \hline
 Conv 3x3x64, ReLU\\
 Conv 3x3x32, ReLU\\
 Max Pool 2x2\\
 Conv 3x3x3, ReLU\\
 Conv 3x3x32, ReLU\\
 Upsampling 2x2\\
 Conv 3x3x64, ReLU\\
 Conv 3x3x64, Sigmoid\\
 \hline
 \end{tabular}
 \end{center}
  \label{table:cifar_dae_arch}
 \end{table}

For the first two datasets, the compression autoencoder has architecture FC-784-81-784. It reduces the input image to 81 dimensions from 784 dimensions, which can also be interpreted as reducing to 9x9 dimensions from 28x28 dimensions. Hence, the size of the input layer of the modified victim CNN is 9x9. This autoencoder was trained with the Adam optimizer with a learning rate of 0.001 and exponential decay rates of 0.9 and 0.999 for the first and second gradient moments, respectively, MSE loss, 100 epochs and a batch size of 500. It has ReLU activation in all layers except the last layer, which has sigmoid activation. The compression autoencoder for the CIFAR-10 dataset has the same architecture as the DAE, and was trained with the same number of epochs, batch size, loss function and optimizer. The difference in its training was that both its input and target output were clean data. It compresses the data from 32x32x3 dimensions to 16x16x3 dimensions, which is the size of the input layer of the modified victim CNN. 

All attacks were generated using the TensorFlow Cleverhans library \cite{cleverhans} as untargeted attacks. The hyperparameters of all three simulated attacks were adjusted such that the attacks are reasonably imperceptible as well as effective in reducing accuracies at the same time. For the PGD attack, the $l_2$ norm of the total distortion was limited by 0.25, the $l_2$ norm of the distortion at each iteration was limited by 0.01, and the attack was performed for 60 iterations. For the Deepfool attack, the maximum number of iterations was set to 50. The CW attack was generated with 5 binary search steps, a maximum of 400 iterations, a learning rate of 0.01, a batch size of 8, an initial constant of 0.01, and the abort\_early parameter was set to True. All other parameters were left at their default values.

It is reasonable to assume that the adversary will be able to make effective guesses about which defense the defender is using. In that case, the attacker will try to generate an attack against the entire victim system, which consists of the defense and the victim network. Such an attack is called an \textit{adaptive attack}, since the adversary adapts their attack to their best guess of the defense. To implement such attacks, instead of simply using gradients of the network, the attacker prepends the preprocessing defense to the network, and uses the gradients of that victim system as a whole to generate the attack. To generate an adaptive attack against multiple defenses, gradients of each network to be attacked are found separately, and are then averaged. Since we have considered a scenario transparent to the attacker, which is similar to a \textit{white box} scenario, all trained classifiers and defenses are accessible to the attacker, and the attacker uses knowledge of these trained classifiers and defenses to generate the attack.

\section{Results} \label{sec: results}

\subsection{Effect of Incorrect Defense Assumptions}\label{subsec: numerical_result}

\begin{table*}
  \caption{Accuracies and replacement-induced robustness for the CIFAR-10 dataset}
  \centering
  
  \begin{tabular}{|m{2em}|m{5em}|m{4em}|m{2em}|m{3em}|m{3em}|m{2em}||m{4em}|m{2em}|m{3em}|m{3em}|m{2em}||p{6em}|}
    \hline
     & &\multicolumn{5}{|c|}{\underline{\textbf{Accuracy}}}  & \multicolumn{5}{|c|}{\underline{\textbf{$r_e(d)$}}} & \underline{\textbf{\makecell{Lower bound \\on $r(d)$}}}\\ \hline
     
     & &\multicolumn{5}{|c|}{\underline{\textbf{$e$}}}  & \multicolumn{5}{|c|}{\underline{\textbf{$e$}}} &\\ \hline

    & &\textbf{None (no defense)}&\textbf{DAE}&\textbf{Cascade}&\textbf{Hidden Layer}&\textbf{AE} & \textbf{None (no defense)}&\textbf{DAE}&\textbf{Cascade}&\textbf{Hidden Layer}&\textbf{AE} & \\ \hline
&\textbf{No attack}&90.44&88.59&84.66&83.3&84.66& && && &\\ \hline
\multicolumn{13}{|c|}{\underline{PGD attack}} \\ \hline

\multirow{5}{1 pt}{\underline{$d$}} & \textbf{None (no defense)} &
\textcolor{blue}{\textbf{39.47}}&85.92&83.89&82.15&83.71& &94.76&98.49&97.74&98.14&98.49\\ \cline{2-13}
&\textbf{DAE}&86.73&\textcolor{blue}{\textbf{57}}&82.01&79.65&83.04&88.26& &91.61&88.45&94.87&94.87\\ \cline{2-13}
&\textbf{Cascade}&89.69&86.46&\textcolor{blue}{\textbf{61.97}}&78.67&73.65&96.69&90.61& &79.59&51.48&96.69\\ \cline{2-13}
&\textbf{Hidden Layer}&89.66&86.41&81.79&\textcolor{blue}{\textbf{55.29}}&82.65&97.22&92.22&89.75& &92.82&97.22\\ \cline{2-13}
&\textbf{AE}&89.43&87.57&76.16&80.52&\textcolor{blue}{\textbf{57.3}}&96.31&96.27&68.93&89.84& &96.31\\ \hline

\multicolumn{13}{|c|}{\underline{CW attack}}\\ \hline
\multirow{5}{1 pt}{\underline{$d$}} & \textbf{None (no defense)} &
\textcolor{blue}{\textbf{6.29}}&87.8&84.27&82.84&84.29& &99.06&99.54&99.45&99.56&99.56\\ \cline{2-13}
&\textbf{DAE}&89.83&\textcolor{blue}{\textbf{7.35}}&83.03&80.57&83.82&99.25& &97.99&96.64&98.97&99.25\\ \cline{2-13}
&\textbf{Cascade}&90.03&86.79&\textcolor{blue}{\textbf{9.01}}&79.82&82.59&99.46&97.62& &95.4&97.26&99.46\\ \cline{2-13}
&\textbf{Hidden Layer}&90.06&86.81&82.91&\textcolor{blue}{\textbf{9.77}}&83.81&99.48&97.58&97.62& &98.84&99.48\\ \cline{2-13}
&\textbf{AE}&90.03&87.96&83.63&81.7&\textcolor{blue}{\textbf{8.87}}&99.46&99.17&98.64&97.89& &99.46\\ \hline
\multicolumn{13}{|c|}{\underline{DF attack}}\\ \hline
\multirow{5}{1 pt}{\underline{$d$}} & \textbf{None (no defense)} &
\textcolor{blue}{\textbf{6.3}}&87.71&84.23&82.67&84.25& &98.95&99.49&99.25&99.51&99.51\\ \cline{2-13}
&\textbf{DAE}&89.59&\textcolor{blue}{\textbf{7.35}}&82.55&80.01&83.43&98.95& &97.4&95.95&98.49&98.95\\ \cline{2-13}
&\textbf{Cascade}&89.98&86.33&\textcolor{blue}{\textbf{8.72}}&78.83&80.84&99.39&97.02& &94.11&94.97&99.39\\ \cline{2-13}
&\textbf{Hidden Layer}&89.91&86.26&82.66&\textcolor{blue}{\textbf{9.58}}&83.4&99.28&96.84&97.29& &98.29&99.28\\ \cline{2-13}
&\textbf{AE}&89.96&87.75&83.35&81.29&\textcolor{blue}{\textbf{8.7}}&99.37&98.89&98.28&97.35& &99.37\\ \hline
\end{tabular}
\label{table: cifar_results}
\end{table*}

\begin{table*}
  \caption{Accuracies and replacement-induced robustness for the Fashion-MNIST dataset with the PGD attack}
  \centering
  \begin{tabular}{|>{\centering\arraybackslash}m{11em}|m{4em}|m{2em}|m{3em}|m{3em}|m{2em}||m{4em}|m{3.5em}|m{3.5em}|m{3.5em}|m{3em}||p{4.5em}|}
    \hline
    \underline{\textbf{\makecell{Defenses attacked}}} & \multicolumn{5}{|c|}{\underline{\textbf{Accuracy}}} & \multicolumn{5}{|c|}{\underline{\textbf{$r_e(D)$}}} & \underline{\textbf{\makecell{Lower limit \\on $r(D)$}}}\\
    \hline
    &\multicolumn{5}{|c|}{\underline{$e$}}&\multicolumn{5}{|c|}{\underline{$e$}}&\\ \hline
    \textbf{$D$}&\textbf{None (no defense)}&\textbf{DAE}&\textbf{Cascade}&\textbf{Hidden Layer}&\textbf{AE} & \textbf{None (no defense)}&\textbf{DAE}&\textbf{Cascade}&\textbf{Hidden Layer}&\textbf{AE} &\\ \hline
\{\}&91.06&83.57&86.12&87.22&87.78& && && &\\ \hline
\textbf{\{None\}}&\textcolor{blue}{\textbf{55.85}}&81.65&85.83&86.9&86.84& &94.55&99.18&99.09&97.33&99.18\\ \hline
\textbf{\{DAE\}}&90.35&\textcolor{blue}{\textbf{63.46}}&84.05&85.22&86.86&96.47& &89.71&90.05&95.43&96.47\\ \hline
\textbf{\{Cascade\}}&90.79&78.69&\textcolor{blue}{\textbf{73.72}}&83.58&85.21&97.82&60.65& &70.65&79.27&97.82\\ \hline
\textbf{\{HL\}}&90.88&78.98&83.05&\textcolor{blue}{\textbf{70.26}}&86.23&98.94&72.94&81.9& &90.86&98.94\\ \hline
\textbf{\{AE\}}&90.01&80.99&80.9&85&\textcolor{blue}{\textbf{77.46}}&89.83&75&49.42&78.49& &89.83\\ \hline

\textbf{\{None, DAE\}}&\textcolor{blue}{\textbf{80}}&\textcolor{blue}{\textbf{69.89}}&79.85&85.31&86.16& &&49.31&84.56&86.9&86.9\\ \hline
\textbf{\{None, Cascade\}}&\textcolor{blue}{\textbf{80.05}}&70.29&\textcolor{blue}{\textbf{84.22}}&78.4&86.63& &-105.73& &-36.64&82.18&82.18\\ \hline
\textbf{\{None, HL\}}&\textcolor{blue}{\textbf{78.9}}&70.06&83.45&\textcolor{blue}{\textbf{85.63}}&82.84& &-96.51&61.16& &28.15&61.16\\ \hline
\textbf{\{None, AE\}}&\textcolor{blue}{\textbf{79.51}}&79.4&78.54&76.87&\textcolor{blue}{\textbf{85.84}}& &38.18&-12.38&-53.45& &38.18\\ \hline
\textbf{\{DAE, Cascade\}}&77.17&\textcolor{blue}{\textbf{79.9}}&\textcolor{blue}{\textbf{77.43}}&85.13&82.02&-124.76& &&66.18&6.8&66.18\\ \hline
\textbf{\{DAE, HL\}}&78.12&\textcolor{blue}{\textbf{79.92}}&82.53&\textcolor{blue}{\textbf{76.49}}&82.23&-79.97& &50.07& &22.81&50.07\\ \hline
\textbf{\{DAE, AE\}}&90.58&\textcolor{blue}{\textbf{69.22}}&79.51&77.86&\textcolor{blue}{\textbf{86.02}}&94.04& &17.94&-16.2& &94.04\\ \hline
\textbf{\{Cascade, HL\}}&90.39&68.34&\textcolor{blue}{\textbf{78.68}}&\textcolor{blue}{\textbf{84.45}}&83.09&86.88&-198.33& &&8.13&86.88\\ \hline
\textbf{\{Cascade, AE\}}&90.44&69.2&\textcolor{blue}{\textbf{82.43}}&77.69&\textcolor{blue}{\textbf{83.15}}&85.1&-245.43& &-129.09& &85.1\\ \hline
\textbf{\{HL, AE\}}&90.69&78.67&76.47&\textcolor{blue}{\textbf{74.68}}&\textcolor{blue}{\textbf{82.23}}&95.91&45.83&-6.69& &&95.91\\ \hline

\textbf{\{None, DAE, Cascade\}}&\textcolor{blue}{\textbf{74.25}}&\textcolor{blue}{\textbf{67.84}}&\textcolor{blue}{\textbf{85.03}}&85.9&87.01& && &88.22&93.13&93.13\\ \hline
\textbf{\{None, DAE, HL\}}&\textcolor{blue}{\textbf{76.55}}&\textcolor{blue}{\textbf{79.62}}&77.17&\textcolor{blue}{\textbf{85.11}}&85.85& &&-30.53& &71.85&71.85\\ \hline
\textbf{\{None, DAE, AE\}}&\textcolor{blue}{\textbf{76.65}}&\textcolor{blue}{\textbf{79.8}}&84.06&74.87&\textcolor{blue}{\textbf{86.44}}& &&68.34&-89.81& &68.34\\ \hline
\textbf{\{None, Cascade, HL\}}&\textcolor{blue}{\textbf{73.61}}&81.18&\textcolor{blue}{\textbf{82.69}}&\textcolor{blue}{\textbf{86.07}}&80.75& &67.45& &&4.27&67.45\\ \hline
\textbf{\{None, Cascade, AE\}}&\textcolor{blue}{\textbf{90.42}}&67.5&\textcolor{blue}{\textbf{78.65}}&84.67&\textcolor{blue}{\textbf{85.89}}& &-382.1& &23.5& &23.5\\ \hline
\textbf{\{None, HL, AE\}}&\textcolor{blue}{\textbf{90.59}}&68.13&83.58&\textcolor{blue}{\textbf{76.91}}&\textcolor{blue}{\textbf{86.55}}& &-285.68&36.55& &&36.55\\ \hline
\textbf{\{DAE, Cascade, HL\}}&90.18&\textcolor{blue}{\textbf{67.63}}&\textcolor{blue}{\textbf{82.56}}&\textcolor{blue}{\textbf{85.16}}&82.02&87.76& && &19.85&87.76\\ \hline
\textbf{\{DAE, Cascade, AE\}}&90.88&\textcolor{blue}{\textbf{78.68}}&\textcolor{blue}{\textbf{76.78}}&73.53&\textcolor{blue}{\textbf{85.38}}&96.75& &&-146.96& &96.75\\ \hline
\textbf{\{DAE, HL, AE\}}&90.4&\textcolor{blue}{\textbf{79.14}}&74.63&\textcolor{blue}{\textbf{83.92}}&\textcolor{blue}{\textbf{80.4}}&86.9& &-128.13& &&86.9\\ \hline
\textbf{\{Cascade, HL, AE\}}&90.49&79.26&\textcolor{blue}{\textbf{81.26}}&\textcolor{blue}{\textbf{72.73}}&\textcolor{blue}{\textbf{81.27}}&93.39&50& && &93.39 \\ \hline

\textbf{\{None, DAE, Cascade, HL\}}&\textcolor{blue}{\textbf{81.61}}&\textcolor{blue}{\textbf{70.7}}&\textcolor{blue}{\textbf{80.51}}&\textcolor{blue}{\textbf{79.14}}&86.14& && &&81.78&81.78\\ \hline
\textbf{\{None, DAE, Cascade, AE\}}&\textcolor{blue}{\textbf{80.53}}&\textcolor{blue}{\textbf{70.41}}&\textcolor{blue}{\textbf{79.85}}&85.17&\textcolor{blue}{\textbf{83.7}}& && &75.91& &75.91\\ \hline
\textbf{\{None, DAE, HL, AE\}}&\textcolor{blue}{\textbf{80.65}}&\textcolor{blue}{\textbf{70.92}}&83.07&\textcolor{blue}{\textbf{79.02}}&\textcolor{blue}{\textbf{83.57}}& &&65.6& &&65.6\\ \hline
\textbf{\{None, Cascade, HL, AE\}}&\textcolor{blue}{\textbf{79.7}}&79.41&\textcolor{blue}{\textbf{78.26}}&\textcolor{blue}{\textbf{77.51}}&\textcolor{blue}{\textbf{82.83}}& &50.89& && &50.89\\ \hline
\textbf{\{DAE, Cascade, HL, AE\}}&90.51&\textcolor{blue}{\textbf{69.58}}&\textcolor{blue}{\textbf{79.19}}&\textcolor{blue}{\textbf{78.28}}&\textcolor{blue}{\textbf{83.65}}&93.53& && &&93.53\\ \hline

\textbf{\{None, DAE, Cascade, HL, AE\}}&\textcolor{blue}{\textbf{71.28}}&\textcolor{blue}{\textbf{81.8}}&\textcolor{blue}{\textbf{80.07}}&\textcolor{blue}{\textbf{79.64}}&\textcolor{blue}{\textbf{83.96}}& && && &\\ \hline

\end{tabular}
\label{table: fashion_results_pgd}
\end{table*}

\begin{table*}
\caption{Accuracies and replacement-induced robustness for the Fashion-MNIST dataset with the CW attack}
  \centering
\begin{tabular}{|>{\centering\arraybackslash}m{11em}|m{4em}|m{2em}|m{3em}|m{3em}|m{2em}||m{4em}|m{3.5em}|m{3.5em}|m{3.5em}|m{3em}||p{4.5em}|}
    \hline
    \underline{\textbf{\makecell{Defenses attacked}}} & \multicolumn{5}{|c|}{\underline{\textbf{Accuracy}}} & \multicolumn{5}{|c|}{\underline{\textbf{$r_e(D)$}}} & \underline{\textbf{\makecell{Lower limit \\on $r(D)$}}}\\
    \hline
&\multicolumn{5}{|c|}{\underline{$e$}}&\multicolumn{5}{|c|}{\underline{$e$}}&\\ \hline
 &\textbf{None (no defense)}&\textbf{DAE}&\textbf{Cascade}&\textbf{Hidden Layer}&\textbf{AE} & \textbf{None (no defense)}&\textbf{DAE}&\textbf{Cascade}&\textbf{Hidden Layer}&\textbf{AE} &\\ \hline
\textbf{\{\}}&91.06&83.57&86.12&87.22&87.78& && && &\\ \hline    
\textbf{\{None\}}&\textcolor{blue}{\textbf{6.55}}&81.88&85.9&86.97&87.29& &98&99.74&99.7&99.42&99.74\\ \hline
\textbf{\{DAE\}}&90.49&\textcolor{blue}{\textbf{11.09}}&83.51&85.15&86.73&99.21& &96.4&97.14&98.55&99.21\\ \hline
\textbf{\{Cascade\}}&90.83&74.82&\textcolor{blue}{\textbf{9.68}}&82.47&85.73&99.7&88.55& &93.79&97.32&99.7\\ \hline
\textbf{\{HL\}}&90.84&78.78&83.23&\textcolor{blue}{\textbf{8.8}}&86.47&99.72&93.89&96.31& &98.33&99.72\\ \hline
\textbf{\{AE\}}&89.18&78.36&80.07&84.81&\textcolor{blue}{\textbf{8.79}}&97.62&93.4&92.34&96.95& &97.62\\ \hline

\textbf{\{None, DAE\}}&\textcolor{blue}{\textbf{45.82}}&\textcolor{blue}{\textbf{48.07}}&85.05&86.42&86.51& &&97.35&98.02&96.85&98.02\\ \hline
\textbf{\{None, Cascade\}}&\textcolor{blue}{\textbf{79.77}}&82.17&\textcolor{blue}{\textbf{79.75}}&86.87&86.1& &84.14& &96.04&80.97&96.04\\ \hline
\textbf{\{None, HL\}}&\textcolor{blue}{\textbf{75.74}}&81.67&85.42&\textcolor{blue}{\textbf{77.48}}&86.88& &84.84&94.41& &92.82&94.41\\ \hline
\textbf{\{None, AE\}}&\textcolor{blue}{\textbf{81.5}}&82.46&84.15&86.89&\textcolor{blue}{\textbf{80.1}}& &87.12&77.15&96.17& &96.17\\ \hline
\textbf{\{DAE, Cascade\}}&88.76&\textcolor{blue}{\textbf{44.9}}&\textcolor{blue}{\textbf{48.99}}&80.58&81.42&93.93& &&82.48&83.22&93.93\\ \hline
\textbf{\{DAE, HL\}}&87.72&\textcolor{blue}{\textbf{31.09}}&78.41&\textcolor{blue}{\textbf{35.08}}&83.33&93.61& &85.26& &91.49&93.61\\ \hline
\textbf{\{DAE, AE\}}&87.55&\textcolor{blue}{\textbf{50.29}}&69.66&80.52&\textcolor{blue}{\textbf{55.52}}&89.29& &49.77&79.55& &89.29\\ \hline
\textbf{\{Cascade, HL\}}&90.31&72.8&\textcolor{blue}{\textbf{32.78}}&\textcolor{blue}{\textbf{31.06}}&78.19&98.63&80.33& &&82.48&98.63\\ \hline
\textbf{\{Cascade, AE\}}&89.34&73.34&\textcolor{blue}{\textbf{9.9}}&80.45&\textcolor{blue}{\textbf{10.2}}&97.76&86.7& &91.2& &97.76\\ \hline
\textbf{\{HL, AE\}}&89.07&72.63&60.35&\textcolor{blue}{\textbf{30.82}}&\textcolor{blue}{\textbf{32.22}}&96.45&80.46&53.97& &&96.45\\ \hline

\textbf{\{None, DAE, Cascade\}}&\textcolor{blue}{\textbf{79.61}}&\textcolor{blue}{\textbf{74.74}}&\textcolor{blue}{\textbf{77.54}}&86.14&85.47& && &88.77&75.99&88.77\\ \hline
\textbf{\{None, DAE, HL\}}&\textcolor{blue}{\textbf{76.62}}&\textcolor{blue}{\textbf{72.2}}&84.52&\textcolor{blue}{\textbf{76.31}}&86.17& &&86.93& &86.85&86.93\\ \hline
\textbf{\{None, DAE, AE\}}&\textcolor{blue}{\textbf{82.53}}&\textcolor{blue}{\textbf{77.17}}&83.02&86.24&\textcolor{blue}{\textbf{80.83}}& &&57.5&86.56& &86.56\\ \hline
\textbf{\{None, Cascade, HL\}}&\textcolor{blue}{\textbf{83.34}}&81.72&\textcolor{blue}{\textbf{80.94}}&\textcolor{blue}{\textbf{81.96}}&85.8& &69.44& &&67.29&69.44\\ \hline
\textbf{\{None, Cascade, AE\}}&\textcolor{blue}{\textbf{78.91}}&81.47&\textcolor{blue}{\textbf{76.47}}&86.37&\textcolor{blue}{\textbf{77.77}}& &80.19& &91.98& &91.98\\ \hline
\textbf{\{None, HL, AE\}}&\textcolor{blue}{\textbf{84.41}}&82.02&83.75&\textcolor{blue}{\textbf{82.56}}&\textcolor{blue}{\textbf{82.62}}& &71.77&56.83& &&71.77\\ \hline
\textbf{\{DAE, Cascade, HL\}}&88.14&\textcolor{blue}{\textbf{47.89}}&\textcolor{blue}{\textbf{51.13}}&\textcolor{blue}{\textbf{51.74}}&76.97&91.75& && &69.45&91.75\\ \hline
\textbf{\{DAE, Cascade, AE\}}&86.45&\textcolor{blue}{\textbf{40.55}}&\textcolor{blue}{\textbf{43.14}}&76.89&\textcolor{blue}{\textbf{44.71}}&89.28& &&75.99& &89.28\\ \hline
\textbf{\{DAE, HL, AE\}}&87.43&\textcolor{blue}{\textbf{52.78}}&65.77&\textcolor{blue}{\textbf{57.18}}&\textcolor{blue}{\textbf{57.4}}&88.06& &33.07& &&88.06\\ \hline
\textbf{\{Cascade, HL, AE\}}&88.94&69.19&22.91&\textcolor{blue}{\textbf{24.34}}&\textcolor{blue}{\textbf{24.15}}&96.65&77.26& && &96.65\\ \hline

\textbf{\{None, DAE, Cascade, HL\}}&\textcolor{blue}{\textbf{83.24}}&\textcolor{blue}{\textbf{77.18}}&\textcolor{blue}{\textbf{79.69}}&\textcolor{blue}{\textbf{80.69}}&85.01& && &&59.22&59.22\\ \hline
\textbf{\{None, DAE, Cascade, AE\}}&\textcolor{blue}{\textbf{80.47}}&\textcolor{blue}{\textbf{75.13}}&\textcolor{blue}{\textbf{77.43}}&85.64&\textcolor{blue}{\textbf{78.84}}& && &82.76& &82.76\\ \hline
\textbf{\{None, DAE, HL, AE\}}&\textcolor{blue}{\textbf{84.5}}&\textcolor{blue}{\textbf{78.29}}&82.78&\textcolor{blue}{\textbf{81.84}}&\textcolor{blue}{\textbf{82.19}}& &&41.43& &&41.43\\ \hline
\textbf{\{None, Cascade, HL, AE\}}&\textcolor{blue}{\textbf{82.57}}&81.24&\textcolor{blue}{\textbf{79.3}}&\textcolor{blue}{\textbf{80.38}}&\textcolor{blue}{\textbf{80.81}}& &67.99& && &67.99\\ \hline
\textbf{\{DAE, Cascade, HL, AE\}}&86.7&\textcolor{blue}{\textbf{46.2}}&\textcolor{blue}{\textbf{48.02}}&\textcolor{blue}{\textbf{49.12}}&\textcolor{blue}{\textbf{49.57}}&88.51& && &&88.51\\ \hline

\textbf{\{None, DAE, Cascade, HL, AE\}}&\textcolor{blue}{\textbf{82.82}}&\textcolor{blue}{\textbf{76.71}}&\textcolor{blue}{\textbf{79.08}}&\textcolor{blue}{\textbf{80.09}}&\textcolor{blue}{\textbf{80.54}}& && && &\\ \hline

\end{tabular}
\label{table: fashion_results_cw}
\end{table*}

 \begin{table*}
 \caption{Accuracies and replacement-induced robustness for the Fashion-MNIST dataset with the DF attack}
  \centering
\begin{tabular}{|>{\centering\arraybackslash}m{11em}|m{4em}|m{2em}|m{3em}|m{3em}|m{2em}||m{4em}|m{3.5em}|m{3.5em}|m{3.5em}|m{3em}||p{4.5em}|}  
  \hline
\underline{\textbf{\makecell{Defenses attacked}}} & \multicolumn{5}{|c|}{\underline{\textbf{Accuracy}}} & \multicolumn{5}{|c|}{\underline{\textbf{$r_e(D)$}}} & \underline{\textbf{\makecell{Lower limit \\on $r(D)$}}}\\
    \hline
&\multicolumn{5}{|c|}{\underline{$e$}}&\multicolumn{5}{|c|}{\underline{$e$}}&\\ \hline
&\textbf{None (no defense)}&\textbf{DAE}&\textbf{Cascade}&\textbf{Hidden Layer}&\textbf{AE} & \textbf{None (no defense)}&\textbf{DAE}&\textbf{Cascade}&\textbf{Hidden Layer}&\textbf{AE} &\\ \hline
\textbf{\{\}}&91.06&83.57&86.12&87.22&87.78& && && &\\ \hline
\textbf{\{None\}}&\textcolor{blue}{\textbf{6.51}}&81.5&85.85&86.95&87.11& &97.55&99.68&99.68&99.21&99.68\\ \hline
\textbf{\{DAE\}}&90.5&\textcolor{blue}{\textbf{11.21}}&82.31&84.15&86.41&99.23& &94.73&95.76&98.11&99.23\\ \hline
\textbf{\{Cascade\}}&90.74&73.66&\textcolor{blue}{\textbf{9.58}}&81.29&85.61&99.58&87.05& &92.25&97.16&99.58\\ \hline
\textbf{\{HL\}}&90.8&78.23&82.71&\textcolor{blue}{\textbf{8.56}}&86.4&99.67&93.21&95.66& &98.25&99.67\\ \hline
\textbf{\{AE\}}&88.98&78.53&79.48&84.54&\textcolor{blue}{\textbf{8.7}}&97.37&93.63&91.6&96.61& &97.37\\ \hline

\textbf{\{None, DAE\}}&\textcolor{blue}{\textbf{6.48}}&\textcolor{blue}{\textbf{11.1}}&82.65&83.63&85.88& &&95.58&95.43&97.58&97.58\\ \hline
\textbf{\{None, Cascade\}}&\textcolor{blue}{\textbf{24.33}}&73.05&\textcolor{blue}{\textbf{67.98}}&82.71&85.82& &75.21& &89.37&95.38&95.38\\ \hline
\textbf{\{None, HL\}}&\textcolor{blue}{\textbf{7.83}}&75.49&82.42&\textcolor{blue}{\textbf{20.31}}&85.43& &89.24&95.07& &96.87&96.87\\ \hline
\textbf{\{None, AE\}}&\textcolor{blue}{\textbf{31.23}}&77.23&81.53&84.84&\textcolor{blue}{\textbf{72.25}}& &83.17&87.82&93.68& &93.68\\ \hline
\textbf{\{DAE, Cascade\}}&90.25&\textcolor{blue}{\textbf{13.99}}&\textcolor{blue}{\textbf{25.53}}&76.11&84.45&98.76& &&82.93&94.88&98.76\\ \hline
\textbf{\{DAE, HL\}}&90.23&\textcolor{blue}{\textbf{11.11}}&78.69&\textcolor{blue}{\textbf{9.8}}&85.17&98.89& &90.09& &96.52&98.89\\ \hline
\textbf{\{DAE, AE\}}&87.65&\textcolor{blue}{\textbf{13.04}}&72.37&79.09&\textcolor{blue}{\textbf{21}}&95.03& &79.97&88.16& &95.03\\ \hline
\textbf{\{Cascade, HL\}}&90.63&69.22&\textcolor{blue}{\textbf{9.93}}&\textcolor{blue}{\textbf{8.43}}&85.04&99.45&81.48& &&96.46&99.45\\ \hline
\textbf{\{Cascade, AE\}}&89.42&73.26&\textcolor{blue}{\textbf{9.46}}&79.95&\textcolor{blue}{\textbf{8.63}}&97.89&86.77& &90.67& &97.89\\ \hline
\textbf{\{HL, AE\}}&88.52&72.69&71.52&\textcolor{blue}{\textbf{8.63}}&\textcolor{blue}{\textbf{8.76}}&96.78&86.19&81.47& &&96.78\\ \hline

\textbf{\{None, DAE, Cascade\}}&\textcolor{blue}{\textbf{29.22}}&\textcolor{blue}{\textbf{18.49}}&\textcolor{blue}{\textbf{66.76}}&77.98&84.48& && &81.05&93.23&93.23\\ \hline
\textbf{\{None, DAE, HL\}}&\textcolor{blue}{\textbf{10.55}}&\textcolor{blue}{\textbf{11.95}}&79.21&\textcolor{blue}{\textbf{33.55}}&84.48& &&89.93& &95.19&95.19\\ \hline
\textbf{\{None, DAE, AE\}}&\textcolor{blue}{\textbf{32.45}}&\textcolor{blue}{\textbf{21.59}}&76.77&80.08&\textcolor{blue}{\textbf{68.74}}& &&79.91&84.66& &84.66\\ \hline
\textbf{\{None, Cascade, HL\}}&\textcolor{blue}{\textbf{29.66}}&68.21&\textcolor{blue}{\textbf{59.32}}&\textcolor{blue}{\textbf{14.55}}&84.88& &71.36& &&94.59&94.59\\ \hline
\textbf{\{None, Cascade, AE\}}&\textcolor{blue}{\textbf{34.86}}&71.63&\textcolor{blue}{\textbf{43.3}}&80.69&\textcolor{blue}{\textbf{39.72}}& &75.65& &86.68& &86.68\\ \hline
\textbf{\{None, HL, AE\}}&\textcolor{blue}{\textbf{32.68}}&71.15&74.86&\textcolor{blue}{\textbf{13.35}}&\textcolor{blue}{\textbf{63.57}}& &76.19&78.41& &&78.41\\ \hline
\textbf{\{DAE, Cascade, HL\}}&90.06&\textcolor{blue}{\textbf{14.72}}&\textcolor{blue}{\textbf{32.21}}&\textcolor{blue}{\textbf{12.01}}&84.17&98.48& && &94.53&98.48\\ \hline
\textbf{\{DAE, Cascade, AE\}}&88.79&\textcolor{blue}{\textbf{15.33}}&\textcolor{blue}{\textbf{19.43}}&74.12&\textcolor{blue}{\textbf{21.98}}&96.61& &&80.42& &96.61\\ \hline
\textbf{\{DAE, HL, AE\}}&87.66&\textcolor{blue}{\textbf{13.73}}&66.82&\textcolor{blue}{\textbf{9.8}}&\textcolor{blue}{\textbf{28.03}}&95.07& &72.03& &&95.07\\ \hline
\textbf{\{Cascade, HL, AE\}}&89.32&68.87&\textcolor{blue}{\textbf{11.45}}&\textcolor{blue}{\textbf{8.76}}&\textcolor{blue}{\textbf{8.75}}&97.75&81& && &97.75\\ \hline

\textbf{\{None, DAE, Cascade, HL\}}&\textcolor{blue}{\textbf{29.23}}&\textcolor{blue}{\textbf{17.33}}&\textcolor{blue}{\textbf{57.37}}&\textcolor{blue}{\textbf{19.87}}&83.44& && &&92.26&92.26\\ \hline
\textbf{\{None, DAE, Cascade, AE\}}&\textcolor{blue}{\textbf{31.69}}&\textcolor{blue}{\textbf{17.63}}&\textcolor{blue}{\textbf{38.29}}&75.52&\textcolor{blue}{\textbf{43.01}}& && &78.52& &78.52\\ \hline
\textbf{\{None, DAE, HL, AE\}}&\textcolor{blue}{\textbf{31.28}}&\textcolor{blue}{\textbf{20.83}}&71.35&\textcolor{blue}{\textbf{17.46}}&\textcolor{blue}{\textbf{61.74}}& &&72.94& &&72.94\\ \hline
\textbf{\{None, Cascade, HL, AE\}}&\textcolor{blue}{\textbf{34.77}}&66.97&\textcolor{blue}{\textbf{38.82}}&\textcolor{blue}{\textbf{13.72}}&\textcolor{blue}{\textbf{38.01}}& &70.73& && &70.73\\ \hline
\textbf{\{DAE, Cascade, HL, AE\}}&88.71&\textcolor{blue}{\textbf{15.99}}&\textcolor{blue}{\textbf{22.67}}&\textcolor{blue}{\textbf{11.01}}&\textcolor{blue}{\textbf{26.83}}&96.5& && &&96.5\\ \hline

\textbf{\{None, DAE, Cascade, HL, AE\}}&\textcolor{blue}{\textbf{30.5}}&\textcolor{blue}{\textbf{17.42}}&\textcolor{blue}{\textbf{35.64}}&\textcolor{blue}{\textbf{15.82}}&\textcolor{blue}{\textbf{40.86}}& && && &\\ \hline

\end{tabular}
\label{table: fashion_results_df}
\end{table*}

\begin{table*}
  \caption{Accuracies and replacement-induced robustness for the MNIST-Digit dataset}
  \centering
  \begin{tabular}{|m{2em}|m{5em}|m{4em}|m{2em}|m{3em}|m{3em}|m{2em}||m{4em}|m{2em}|m{3em}|m{3em}|m{2em}||p{6em}|}
    \hline
     & &\multicolumn{5}{|c|}{\underline{\textbf{Accuracy}}}  & \multicolumn{5}{|c|}{\underline{\textbf{$r_e(d)$}}} & \underline{\textbf{\makecell{Lower bound \\on $r(d)$}}}\\ \hline
     
     & &\multicolumn{5}{|c|}{\underline{\textbf{$e$}}}  & \multicolumn{5}{|c|}{\underline{\textbf{$e$}}} &\\ \hline

    & &\textbf{None (no defense)}&\textbf{DAE}&\textbf{Cascade}&\textbf{Hidden Layer}&\textbf{AE} & \textbf{None (no defense)}&\textbf{DAE}&\textbf{Cascade}&\textbf{Hidden Layer}&\textbf{AE} & \\ \hline
&\textbf{No attack} &98.66&98.16&96.54&96.6&96.65& && && &\\ \hline
\multicolumn{13}{|c|}{\underline{PGD attack}} \\ \hline
\multirow{5}{1 pt}{\underline{$d$}} & \textbf{None (no defense)} & \textcolor{blue}{\textbf{97.42}}&97.95&96.46&96.47&96.16& &83.06&93.55&89.52&60.48&93.55\\ \cline{2-13}
&\textbf{DAE}&98.62&\textcolor{blue}{\textbf{53.46}}&74.37&63.63&96.51&99.91& &50.4&26.24&99.69&99.91\\ \cline{2-13}
&\textbf{Cascade}&98.64&79.11&\textcolor{blue}{\textbf{29.07}}&45.51&96.51&99.97&71.77& &24.28&99.79&99.97\\ \cline{2-13}
&\textbf{Hidden Layer}&98.62&85.63&71.81&\textcolor{blue}{\textbf{16.33}}&96.51&99.95&84.39&69.19& &99.83&99.95\\ \cline{2-13}
&\textbf{AE}&98.54&98.1&95.83&95.88&\textcolor{blue}{\textbf{91.87}}&97.49&98.74&85.15&84.94& &98.74\\ \hline
\multicolumn{13}{|c|}{\underline{CW attack}} \\ \hline
\multirow{5}{1 pt}{\underline{$d$}} & \textbf{None (no defense)} &\textcolor{blue}{\textbf{1.08}}&97.62&96.09&95.74&94.58& &99.45&99.54&99.12&97.88&99.54\\ \cline{2-13}
&\textbf{DAE}&98.65&\textcolor{blue}{\textbf{1.42}}&75.96&56.39&96.59&99.99& &78.73&58.43&99.94&99.99\\ \cline{2-13}
&\textbf{Cascade}&98.65&93.43&\textcolor{blue}{\textbf{2.32}}&62.62&96.65&99.99&94.98& &63.94&100&100\\ \cline{2-13}
&\textbf{Hidden Layer}&98.66&97.29&93.93&\textcolor{blue}{\textbf{2.41}}&96.64&100&99.08&97.23& &99.99&100\\ \cline{2-13}
&\textbf{AE}&98.46&97.94&96.12&94.83&\textcolor{blue}{\textbf{2.37}}&99.79&99.77&99.55&98.12& &99.79\\ \hline
\multicolumn{13}{|c|}{\underline{DF attack}} \\ \hline
\multirow{5}{1 pt}{\underline{$d$}} & \textbf{None (no defense)} &\textcolor{blue}{\textbf{1.09}}&97.51&96.01&95.8&93.97& &99.33&99.46&99.18&97.25&99.46\\ \cline{2-13}
&\textbf{DAE}&98.64&\textcolor{blue}{\textbf{1.4}}&71.42&48.04&96.57&99.98& &74.04&49.81&99.92&99.98\\ \cline{2-13}
&\textbf{Cascade}&98.65&91.72&\textcolor{blue}{\textbf{2.27}}&58.08&96.64&99.99&93.17& &59.14&99.99&99.99\\ \cline{2-13}
&\textbf{Hidden Layer}&98.66&97.33&93.73&\textcolor{blue}{\textbf{2.32}}&96.62&100&99.12&97.02& &99.97&100\\ \cline{2-13}
&\textbf{AE}&98.45&98&96.09&94.71&\textcolor{blue}{\textbf{2.39}}&99.78&99.83&99.52&97.99& &99.83\\ \hline
\end{tabular}
\label{table: digit_results}
\end{table*}

Apart from considering our two proposed defenses, we also considered a simple DAE defense, a simple autoencoder defense, and the case when no defense is used. A \textit{DAE defense} means that the test data is denoised using a DAE before being input into the unmodified victim classifier. An \textit{autoencoder defense} means that the test data was compressed using the compression autoencoder before being input into an AE-reduced classifier. This is abbreviated as the \textit{AE defense}. We recorded the performance of each of these defenses in the presence of no attack, in the presence of a non-adaptive attack (also referred to as a naive attack), and in the presence of adaptive attacks to circumvent one defense at a time. Tables \ref{table: cifar_results} and \ref{table: digit_results} show the results of our experiments for the CIFAR-10 dataset and the MNIST-Digit dataset, respectively, using the PGD, CW and DF attacks. Further, for the Fashion-MNIST dataset, we also considered adaptive attacks to circumvent multiple defenses at the same time, as shown in Tables \ref{table: fashion_results_pgd}, \ref{table: fashion_results_cw} and \ref{table: fashion_results_df}.

Each table showing the results is vertically divided into three parts, where each part is separated by a double vertical line. The first vertical part, consisting of five columns, displays accuracies. The remaining parts show calculations for the replacement-induced robustness, which is introduced in Section \ref{subsec: metric}. 
As labeled, the first row corresponds to a case when there is no attack. Each of the remaining rows corresponds to a case when there is an attack, targeting zero, one or more defenses. The row labeled `None (no defense)' corresponds to a case when the attack is naive, i.e., it does not attempt to circumvent the defense, but attempts to attack the original victim network. The remaining rows illustrate scenarios when one or more defenses are targeted. Each column corresponds to a particular defense used by the defender. In the first vertical part of a given row, the entries that are in bold and blue show accuracies when the defense assumed by the attacker is indeed used.

There are ${5 \choose 1}=5$ ways to attack one defense, ${5 \choose 2}=10$ ways to attack two defenses, ${5 \choose 3}=10$ ways to attack three defenses, ${5 \choose 4}=5$ ways to attack four defenses, and ${5 \choose 5}=1$ way to attack all five defenses. Additionally, there is ${5 \choose 0}=1$ way to attack none of the defenses. We have shown each of these $2^5=32$ scenarios for the Fashion-MNIST dataset in Tables \ref{table: fashion_results_pgd}, \ref{table: fashion_results_cw} and \ref{table: fashion_results_df} for the PGD, CW and DF attacks, respectively.


We observe that when the attacker uses an attack to circumvent one specific defense, the accuracy dips quite low if that defense is indeed used. 
However, when there is a mismatch between the defense assumed by the attacker and the defense actually used by the defender, the accuracy may be significantly higher, as shown by all the accuracies that are not in bold and blue.
This means that if the attacker cannot correctly guess which defense is being used, an adaptive attack becomes very weak. \textbf{Such an attack is likely to be especially weak when the defender chooses to not use a defense at all}, as shown by the first column of the result tables in the vertical part showing accuracies. This loophole can be exploited by the defender through \textbf{having a repertoire containing multiple defenses, and choosing one (or none) randomly at test time}, so that the knowledge about which defense will be used cannot be made available in advance.
If the attacker tries to overcome multiple defenses at the same time, then the attack becomes weaker, as seen in Tables \ref{table: fashion_results_pgd}, \ref{table: fashion_results_cw} and \ref{table: fashion_results_df}.
Thus, we observe that while our considered defenses are vulnerable to adaptive attacks in the presence of perfect knowledge about the deployed defense, such attacks can be overcome by maintaining uncertainty about which defense will be used at test time, as well as keeping the possibility of having no defense. Further, we identify that in the absence of an attack, using a defense does hurt the accuracy slightly; a manifestation of the robustness-performance tradeoff.

\subsection{Metric to Evaluate Defenses} \label{subsec: metric}

For a given defense $d$, let us refer to the adaptive attack designed to circumvent the defense as $a(d)$, termed as \textit{adaption against d}. Since it is possible to generate an adaptive attack to circumvent multiple defenses simultaneously, we also consider $a(D)$ for a set of defenses $D$.
Let us now define $s(a(d))$, termed as \textit{success of a(d)}, as the decrease in accuracy when the attack is $a(d)$ and the defense used is $d$, from the accuracy when there is no attack and the defense used is $d$. For a set of defenses $D$, we define $s(a(D))$ as the decrease in the average of accuracies over $d \in D$ when the attack is $a(D)$ and the defense used is $d$, from the average of accuracies over $d \in D$ when there is no attack and the defense used is $d$. We may view $s(a(d))$ as a special case of $s(a(D))$, where $D$ has a single element, i.e., $D = \{d\}$. Hence, for easier generalization, we continue our discussion by considering a set of defenses $D = \{d_1, d_2, ..., d_i, ...\}$ attacked by the attacker.
The value of $s(a(D))$ indicates how successful $a(D)$ was in decreasing the accuracy, assuming that $d \in D$ is used as a defense. However, the defender is free to use a defense $e \notin D$, which motivates the need to measure the success of the attack when such a defense $e$ is used. Let us define $s(a(D)|e)$, termed as \textit{success of a(D) given e}, as the decrease in accuracy when the attack is $a(D)$ and the actual defense used is $e$, from when there is no attack and the defense used is $e$. Here, $e$ may or may not belong to $D$.

To determine if $a(D)$ is as successful in circumventing a defense $e \notin D$ as it is in circumventing other defenses $d \in D$, we need to compare $s(a(D)|e)$ and $s(a(D))$. Hence, for defense $e \notin D$, we define $r_e(D)$, termed as the \textit{replacement-induced robustness of D to a(D) via e}, as the percentage by which $s(a(D)|e)$ is less than $s(a(D))$.
\begin{equation}
    r_e(D) = \bigg(\frac{s(a(D)) - s(a(D)|e)}{s(a(D))}\bigg)100.
    \label{eq: r_e(d)}
\end{equation}
For $e \in D$, $r_e(D)$ is undefined. For a single defense $d$, we define the \textit{replacement-induced robustness of $d$ to $a(d)$ via $e$} as being equivalent to the \textit{replacement-induced robustness of $\{d\}$ to $a\left(\{d\}\right)$ via $e$}, i.e., $r_e(d) \equiv r_e\left(\{d\}\right)$.

The replacement-induced robustness metric measures the loss incurred by the attacker if the defender decides to use a defense $e\notin D$ instead of a targeted defense $d \in D$. If we want to evaluate how vulnerable the performance of $a(D)$ is to changes in the actual defense used, we can look at the maximum value for $r_e(D)$ over all possible $e$. \textbf{Let $E$ be the set of all possible defenses $e \notin D$ that deliver acceptable performance in presence of no attack}; possibly including the strategy of having \textit{no defense}. Then, we define $r(D)$, termed as \textit{universal replacement-induced robustness of D to a(D)}, as the aforementioned maximum:

\begin{equation}
    r(D) 
    = \max_{e \in E} (r_e(D)).
    \label{eq: r(d)}
\end{equation}
\begin{figure}[htb]
  \centering 
  \subfigure[Orders when the classifier is chosen as the first component to be trained.]{\label{subfig: tree1}  \includegraphics[width=0.38\textwidth]{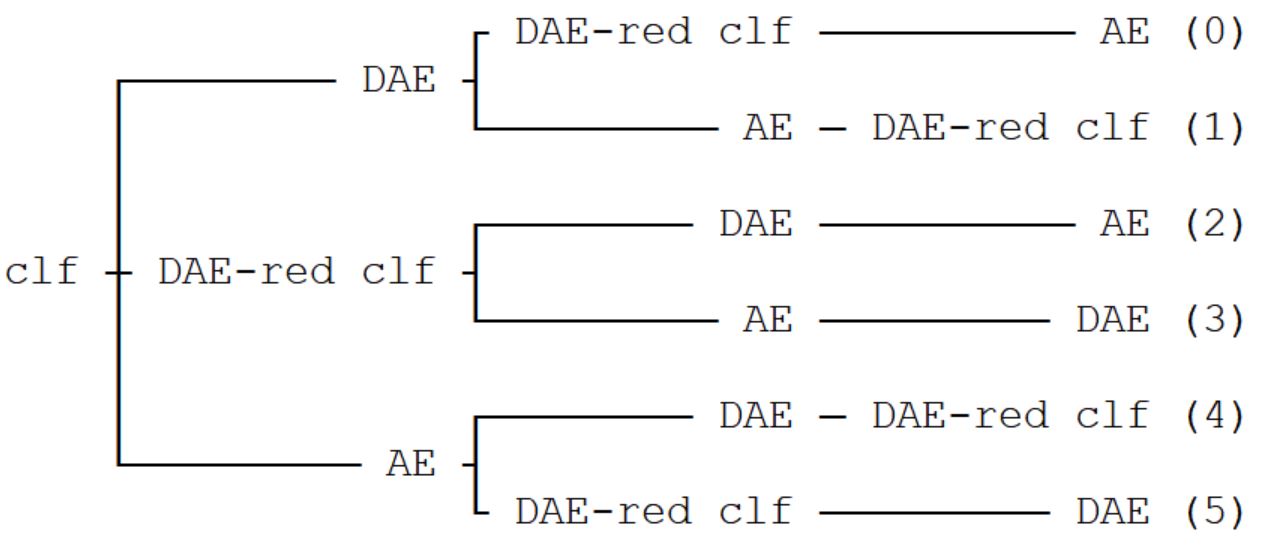}}%
  \hskip 0.5truein
  \subfigure[Orders when the DAE is chosen as the first component to be trained.]{\label{subfig: tree2}\includegraphics[width=0.38\textwidth]{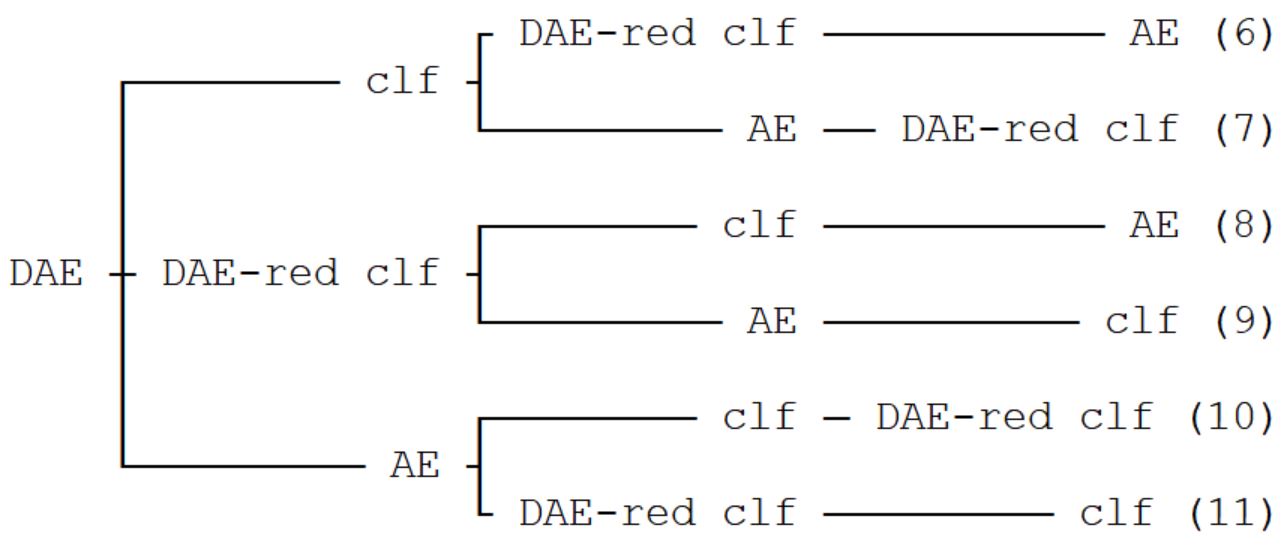}}
  \hskip 0.5truein
  \subfigure[Orders when the AE is chosen as the first component to be trained.]{\label{subfig: tree3}\includegraphics[width=0.38\textwidth]{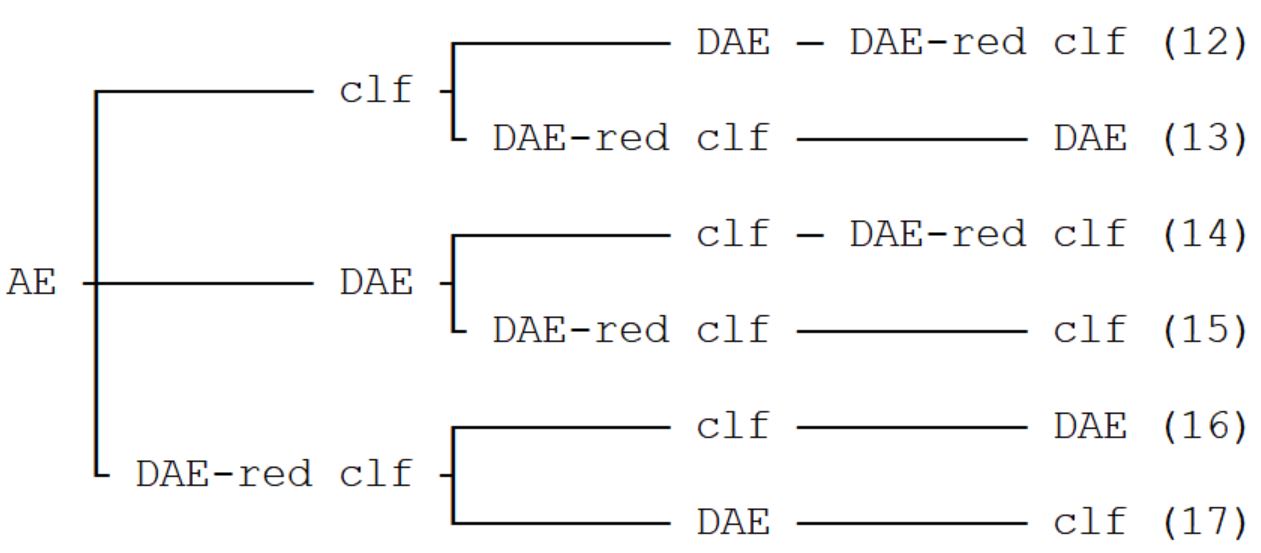}}
  \caption{Orders in which different components of the defense infrastructure can be trained.}
  \label{fig: tree}
\end{figure}

For a single defense $d$, we define $r(d) \equiv r(\{d\})$. A positive value of the universal replacement-induced robustness of $D$ to $a(D)$ tells us that there exists some defense $e$, through which the defender can diminish the success of the attacker. Further, the value of this metric itself tells us the percentage by which the success of the attacker can be reduced.
It may be hard to consider all $e \in E$, and then calculate $s(a(D)|e)$ for each. However, we can typically find a lower bound on $r(D)$, by evaluating $r_e(D)$ for practically significant choices of $e$, and taking the maximum as the lower bound. 

In Tables \ref{table: cifar_results}-\ref{table: digit_results}, the second vertical part shows the replacement-induced robustness of $D$ to $a(D)$ via $e$. Here, $e$, which is the actual defense used, is indicated by the column, and $D$, which refers to the set of defenses attacked, is indicated by the row. For the CIFAR-10 dataset and the MNIST-Digit dataset whose results can be found in Tables \ref{table: cifar_results} and \ref{table: digit_results}, respectively, we only show the case when $D$ consists of at most one defense, i.e., $D=\{\}$ or $D=\{d\}$. Note that in the columns showing $r_e(D)$, we have left the corresponding cell blank when $e \in D$, since $r_e(D)$ is undefined for such cases.
The third vertical part of each of these tables shows a lower bound on the universal replacement-induced robustness of $D$ to $a(D)$.
We observe that $r(D)$ is indeed typically very high, showing that using a defense that the attacker was unprepared for can significantly reduce the effectiveness of the attack. Moreover, we notice that $r_e(D)$ is positive in most cases, signifying that an arbitrary defense choice is likely to reduce the success of the attacker. 
The only exceptions are a few cases when the network to classify the Fashion-MNIST dataset is attacked with the PGD attack. However, all of these cases correspond to attacks with mild successes. Moreover, even in these cases, $r(D)$ stays positive with a significant magnitude, indicating that there exist defenses which can be used to decrease the success of the attacker significantly.

The universal replacement-induced robustness of a defense $d$ or a set of defenses $D$ is an attribute that indeed creates a setback for the attacker when it tries to circumvent the defense(s). For sets of defenses $D$ having large values of this attribute, it becomes difficult for the attacker to generate an attack which cannot be circumvented, while also maintaining the effectiveness of the attack.

\subsection{Guide to Building Defense Infrastructure} \label{subsec: guide}

\begin{table}
  \caption{Training times for networks for the Fashion-MNIST dataset}
  \centering
  \begin{tabular}{|m{17em}|m{6em}|}
    \hline
     \textbf{Network}&\textbf{Time (second)}\\ \hline
     Classifier with full input dimension&42.13\\ \hline
    Classifier with reduced input dimension&15.46 \\ \hline
    DAE&300.0 \\ \hline
    AE&60.0 \\ \hline
    \end{tabular}
\label{table: training_times}
\end{table}

\begin{table*}
  \caption{Orders of training components of the defense infrastructure}
  \centering
  \begin{tabular}{|m{12em}|m{9em}|m{9em}|m{9em}|m{9em}|}
    \hline
     & \multicolumn{4}{|c|}{$i$}\\ \hline
     &1&2&3&4\\ \hline \hline
     
\multicolumn{5}{|c|}{\textbf{Order 1}}\\ \hline
     
Component trained at step $i$&clf&DAE&AE&DAE-red clf \\ \hline
Training time for step $i$ (s)&42.13&342.13&417.59&433.05\\ \hline
Defenses added after step $i$&None&DAE&AE, cascade&HL\\ \hline
Avg accuracy after step $i$ (\%)&6.55&46.95&77.97&79.85\\ \hline \hline
\multicolumn{5}{|c|}{\textbf{Order 4}}\\ \hline
Component trained at step $i$&clf&AE&DAE&DAE-red clf\\ \hline
Training time after step $i$ (s)&42.13&117.59&417.59&433.05\\ \hline
Defenses added after step $i$&None&AE&DAE, cascade&HL\\ \hline
Avg accuracy after step $i$ (\%)&6.55&80.8&77.97&79.85\\ \hline \hline
\multicolumn{5}{|c|}{\textbf{Order 8}}\\ \hline
Component trained at step $i$&DAE&DAE-red clf&clf&AE\\ \hline
Training time after step $i$ (s)&300&315.46&357.59&433.05\\ \hline
Defenses added after step $i$&&HL&None, DAE&AE, cascade\\ \hline
Avg accuracy after step $i$ (\%)&&8.8&75.04&79.85\\ \hline \hline
\multicolumn{5}{|c|}{\textbf{Order 9}}\\ \hline
Component trained at step $i$&DAE&DAE-red clf&AE&clf\\ \hline
Training time after step $i$ (s)&300&315.46&390.92&433.05\\ \hline
Defenses added after step $i$&&HL&AE&None, DAE, cascade\\ \hline
Avg accuracy after step $i$ (\%)&&8.8&31.52&79.85\\ \hline \hline
\multicolumn{5}{|c|}{\textbf{Order 12}}\\ \hline
Component trained at step $i$&AE&clf&DAE&DAE-red clf\\ \hline
Training time after step $i$ (s)&75.46&117.59&417.59&433.05\\ \hline
Defenses added after step $i$&AE&None&DAE, cascade&HL\\ \hline
Avg accuracy after step $i$ (\%)&8.79&80.8&77.97&79.85\\ \hline \hline
\end{tabular}
\label{table: guide}
\end{table*}

We acknowledge that a limitation in computational resources can restrict the capabilities of a defender, who may have only budgeted for the computational resources needed to train the victim network, and may be short on computational resources to train the defense. This motivates us to discuss how the defender can craft their victim system to be as robust as possible given their computational bottleneck. We divide the process of building the repertoire of defenses into small steps, such that completion of each step guarantees the availability of a minimum number of defenses, even if there are not enough resources to complete the remaining steps. While building this guide to construct the repertoire, we consider all reasonable orders in which the components of the victim system can be trained, and for each reasonable order, consider the performance achieved at each step of the building process. Since the performance depends on the dataset, classifiers used and the algorithm used by the attacker, we have narrowed down our analysis to the scenario when our specific classifier described in Section \ref{subsec: impl_details} is used for the Fashion-MNIST classification task, and the attacker chooses to use the Carlini-Wagner attack. Thus, the following discussion is a specific example of the process of constructing a guide for a given classifier, dataset and attack algorithm.

We consider the orders in which the components of the victim system can be trained. The components considered are the classifier, the DAE, the DAE-reduced classifier, the compression autoencoder and the AE-reduced classifier. Among these, the compression autoencoder and the AE-reduced classifier are considered as one component, because one does not provide a useful defense without the other. This component is abbreviated as the `AE' component. Further, the classifier is abbreviated as the `clf' component and the DAE-reduced classifier is abbreviated as the `DAE-red clf' component.
Figure \ref{fig: tree} shows all possible orders in which the different components can be trained. This is shown in the form of trees, where each node corresponds to a component, and the path from the root to a leaf shows one possible order. Each leaf is labelled with an integer in parentheses to help reference that specific order of components. The first, second and third trees show all possible orders when the clf, DAE and AE are respectively chosen as the first component to be trained. Next, we reduce our list of orders to only those that seem reasonable. Order 0 seems unreasonable because after having trained the first two components in that order, it is wiser to train the AE before the DAE-red clf, since training the AE first affords us both the AE and the cascaded defenses, as in Order 1, while training the DAE-red clf first only affords us the HL defense. Orders 2, 3, 5, 13, 16 and 17 are unreasonable because the DAE-red clf cannot be trained before the DAE is. Orders 6 and 7 are eliminated from consideration because they are similar to Orders 0 and 1, since training the DAE before the clf cannot afford us any defenses that training the clf before the DAE cannot. Order 14, being similar to Order 12, is eliminated for the same reason. 
Orders 10 and 11 are not considered because it does not make sense to first train a DAE and then train the AE without training any classifier(s) to make use of the DAE first. Order 15 is also unreasonable because after having trained the first two components, training the DAE-red clf affords us only the option of using the Hidden Layer defense, while training the clf instead would afford us three options, i.e., the DAE defense, the cascaded defense and the possibility of not using a defense. Hence we are left with Orders 1, 4, 8, 9 and 12, which we examine further in Table \ref{table: guide}.

Table \ref{table: guide} is divided into five horizontal sections separated by double lines, each of which describes one of the five orders we have shortlisted. The four columns of the table correspond to the four steps indexed by a variable $i$. Each horizontal section has five rows. The first and second rows mention the numerical identifier of the order and the order of training of the components respectively. The third row lists the cumulative amount of time taken to reach the end of that step, while the fourth row lists the defenses added to our repertoire at the end of each step. These defense names correspond to the defense names used in Tables \ref{table: cifar_results} - \ref{table: digit_results}. The fifth row shows the average accuracy that our repertoire can achieve at the end of each step. This is the average of the accuracies when the attacker attacks all available defenses, and one defense is used at a time. The average is taken over all cases where each case corresponds to one available defense being used. 

Considering the average accuracy at the end of each step, Orders 4 and 12 appear promising. In both of these orders, after the second step completes, which takes place after only 117.59 seconds, we achieve an average accuracy as high as 80.8\%. This is because having the `None' option for a defense available with a second defense makes our repertoire quite strong. This is true especially if the second defense is cascade, HL or AE. Order 1 also achieves the availability of the `None' defense option along with the DAE defense. However, this order is not as desirable since the average accuracy for such a pair of defenses is not as high, and also because the DAE component takes longer to train than the AE component. Orders 8 and 9 are also much less preferred because they present no defense infrastructure available after the first step completes, and are only able to achieve an average accuracy of 8.8\% after the second step completes, while taking a large amount of time to finish that second step. 

\section{Concluding Remarks}

If we consider each of the proposed defenses in a standalone manner, then using the cascaded defense would offer both robustness and computationally cheaper training for the victim network, at the cost of extra computational power needed to train the compression autoencoder. Contrarily, using only the DAE defense does not offer computational reduction for training the victim network. If training the autoencoder for dimensionality reduction requires more computational power than the defender can afford, the proposed hidden layer defense offers comparable robustness while maintaining the advantage of computationally cheaper victim network training.

However, like most other standalone defenses in existing literature, the proposed defenses are vulnerable to adaptive attacks in presence of exact knowledge about the defense being used. Hence, our contribution is best viewed in light of the discovery that an adaptive attack designed to circumvent a specific defense among the proposed, becomes weak under the influence of a different defense, or even if no defense is used. Further, an adaptive attack attempting to simultaneously circumvent multiple defenses becomes weak, and hence, hindering the attacker's ability to effectively use perfect knowledge of the available repertoire of defenses. Capitalizing on this discovery, we have proposed a systematic guide to construct a defense infrastructure that is equipped to remove both predictable and unforeseen perturbations, by smartly using available computational resources. It is important to note that the scope of this discovery may extend beyond the specific defenses considered in this work, and future work will confirm if other defenses also possess replacement-induced robustness to adaptive attacks. If so, then this is a useful step towards finding a generally effective strategy to defend against adaptive adversarial attacks.

\section*{Acknowledgment}
The authors would like to thank Milind Kulkarni of Purdue ECE for fruitful discussions, specially regarding the performance-robustness tradeoff and the use of no defense to circumvent adaptive attacks.
\bibliographystyle{ieeetr}
\bibliography{all}
\end{document}